\documentclass[a4paper,12pt]{amsart}
\usepackage{dsfont}
\usepackage[english]{babel}
\usepackage[utf8]{inputenc}
\usepackage[T1]{fontenc}
\usepackage{hyperref}
\usepackage{graphicx}
\usepackage{amsmath}
\usepackage{amssymb}
\DeclareMathAlphabet{\altmathcal}{OMS}{cmsy}{m}{n}
\usepackage{mathptmx}

\newtheorem{theorem}{Theorem}
\newtheorem{definition}{Definition}[section]
\newtheorem{lemma}{Lemma}

\usepackage{fancyhdr}
\renewcommand{\contentsline}[4]{\csname nuova#1\endcsname{#2}{#3}{#4}}
\newcommand{\nuovasection}[3]{\medskip\hbox to \hsize{\vbox{\advance\hsize by -1cm\baselineskip=9pt\parfillskip=0pt\leftskip=3.5cm\noindent\hskip -2cm #1\leaders\hbox{.}\hfil\hfil\par}$\,$#2\hfil}}
\newcommand{\nuovasubsection}[3]{\smallskip\hbox to \hsize{\vbox{\advance\hsize by -1cm\baselineskip=10pt\parfillskip=0pt\leftskip=4cm\noindent\hskip -2cm #1\leaders\hbox{.}\hfil\hfil\par}$\,$#2\hfil}}

\let\oldmaketitle\maketitle
\renewcommand\maketitle{{\bfseries\boldmath\oldmaketitle}}

\begin{document}
				
   \vspace{-8.0cm}
            \title{On the Integrability of Pfaffian Forms on ${\mathds R}^{n}$}
            \author{Pedro F. da Silva J{\'u}nior}
	\address{Universidade Federal de Pernambuco\hfill\break\indent Caruaru, PE, Brazil}
						\email{pedro.fsilva2@ufpe.br}

\begin{abstract}
This paper details the lesser known conditions on ${\mathds R}^{n}$ for the integrability of pfaffian forms, or 1-forms. Emphasis is given to locality of these conditions, and proofs in some additional detail are provided for theorems due to Clairaut and Carathéodory. Considering the importance of the integrability of pfaffian forms, in particular in mathematical-physics, this paper shows that: there is a hidden content in Carathéodory's theorem in the direction of a global integrability. 
\end{abstract}
				 
				\maketitle
				\thispagestyle{plain}
\vspace{-0.6cm}
\section{Introduction}

Pfaffian forms are a particular case of differential forms: the \textit{1-forms}. The treatment of this subject via differential forms, including the use of exterior algebra and differentiable manifolds more general than ${\mathbb {R}}^{n}$, is beyond the scope of this paper\footnote{To the reader interested in an extensive approach to the subject via differential forms, we suggest consulting the book by Flanders \cite{flanders}, for a applied exposition in physics, or the book by Morita \cite{morita}, for one based in pure mathematics.}. The study of pfaffian forms has theoretical and practical relevance in itself, particularly, in mathematical-physics \cite{antoniou,arens}.

This paper has two main objectives. The first is to present to the reader, in some detail, the lesser known conditions for the integrability of pfaffian forms on ${\mathbb {R}}^{n}$, which have a \textit{local} character and rarely appear in textbooks of differential equations and differential forms. The second is to discuss the possibility and obtainability of a \textit{global} integrability criterion for pfaffian forms.  Frobenius Theorem does not fit the proposed roadmap for this paper and thus will be omitted. Exception is a briefly comment in section \ref{sec:finais}.

Well, first studied by Clairaut, Fontaine, and Euler, according to Katz \cite{katz}, the pfaffian forms were so named in honor of Pfaff, who between 1814 and 1815, treated the subject in greater detail \cite{samelson}. Notable mathematicians later went on to expand on Pfaff's work, most notably Frobenius \cite{frobenius} and Cartan \cite{cartan}. 

A suitable way to define pfaffian forms here is similar to how Morita \cite{morita} does.

\begin{definition}\label{itm:1.1}{\normalfont(Pfaffian form)}. Let be a collection of n independent variables $x_{1},x_{2},...,x_{n}$ in ${\mathbb {R}}^{n}$, and a collection of n functions $F_{i}=F_{i}(x_{1},x_{2},...,x_{n})$ of class $C^{\infty}$ on an open set $B\:{\subseteq}\:{\mathbb {R}}^{n}$. For the $\delta\xi$ objects,
\begin{align*}
\delta{\xi}=\sum_{i=1}^{n}F_{i}(x_{1},x_{2},...,x_{n})dx_{i},
\end{align*}
defined in $B$, with $\delta\xi$ representing the infinitesimal of a certain finite quantity $\xi$ in B, we call them pfaffian forms in n variables.
\end{definition}

Henceforth when we refer to a generic pfaffian form we will be referring to a pfaffian form $\delta\xi$, whose parameters are exemplified from Definition \ref{itm:1.1}. Now, a pfaffian form may, or may not, refer to what we call the differential of a function. When a pfaffian form $\delta\xi$ does not have the functions $F_{i}$ identified as the partial derivatives of a function $\xi=\xi(x_{1},x_{2},... ,x_{n})$ with respect to the respective variables $x_{i}$, ($F_{i}={\partial}{\xi}/{\partial}x_{i}$), then it does not represent the differential of such a function $\xi$ and we call $\delta\xi$ an inexact differential. That is, in this case, the quantity ${\delta}{\xi}$ represents only the infinitesimal of a certain finite quantity ${\xi}$, and ${\xi}$ is not a function of the $n$ independent variables $x_{i}$ in the sense of ${\xi}=\xi(x_{1},x_{2},...,x_{n})$.

Otherwise, if the pfaffian form ${\delta}\xi$ has the functions $F_{i}$ identified as the partial derivatives of a function ${\xi}=\xi(x_{1},x_{2},.... ,x_{n})$ with respect to the respective variables $x_{i}$, ($F_{i}={\partial}{\xi}/{\partial}x_{i}$), then it is the differential of such a function $\xi$, actually existing. Moreover, in this case, we also call $\delta{\xi}$ an exact differential and replace, in the symbolism of its designation, the symbol $\delta$ by the usual symbol $d$, traditionally used to denote the infinitesimal of a quantity that is a usual function.

It turns out that in some cases, even if the pfaffian form $\delta\xi$ is an inexact differential, it can be written as the result of the product of a function ${\mu}={\mu}(x_{1},x_{2},...,x_{n})$ with an exact differential $d\psi$, where $d\psi$ is a function of the $n$ independent variables $x_{1},x_{2},. ...,x_{n}$, that is, $\psi=\psi(x_{1},x_{2},...,x_{n})$. In other words, there being $\mu$, and being $\mu(x_{1},x_{2},...,x_{n})\;{\neq}\;0$, in an open set $A$, where $A\:{\subseteq}\:B$, it follows that the quantity ${\delta}{\xi}/{\mu}$ will be an exact differential $d\psi$ in $A$. When this happens we say that $\delta\xi$ is integrable in $A$, and we call the function ${\mu}^{-1}$ the integrating factor of $\delta\xi$. Furthermore, given the smoothness of functions $F_{i}$ on all $B$, for the discussion of integrability we will also assume that the studied pfaffian forms are always \textit{non-singular} in $B$; i.e., not identically null on all $B$.

\begin{definition}\label{itm:1.2}{\normalfont(Integrable pfaffian form)}. Let $\delta\xi$ be a non-singular pfaffian form. If there exist functions ${\mu}={\mu}(x_{1},x_{2},...,x_{n})$, with $\mu(x_{1},x_{2},...,x_{n})\;{\neq}\;0$, and $\psi=\psi(x_{1},x_{2},...,x_{n})$ such that
\begin{align*}
\delta{\xi}=\mu{d\psi},
\end{align*}
on an open $A\:{\subseteq}\:B$, then $\delta\xi$ is said to be integrable on $A$. Also, the function ${\mu}^{-1}$ is called the integrating factor of $\delta\xi$.
\end{definition}

Naturally, by Definition \ref{itm:1.2}, every $\delta\xi$ such that $\delta\xi=d\xi$ is integrable.
To continue our discussion we need to mention the important situation that occurs on paths in $B$ such that a pfaffian form nullifies, where we get the so-called Pfaff equation associated with that pfaffian form.

\begin{definition}\label{itm:1.3}{\normalfont(Pfaff equation)}. The associated Pfaff equation for the pfaffian form $\delta{\xi}$ is 
\begin{align*}
\delta{\xi}=0.
\end{align*} 
\end{definition}

It is very important to be said that a Pfaff equation \textit{not} says that $\delta{\xi}$ is identically null at $B$; that equation says at which \textit{paths} of $B$ the equation $\delta{\xi}=0$ has solution\footnote{This can be exemplified with some uses of Pfaff equations in physics. In Analytic Mechanics, {\textit{constraints}} are often modeled by a Pfaff equation \cite{papastavridis}, and in Classical Thermodynamics the usual condition for an adiabatic {\textit{infinitesimal process}} $\delta{\altmathcal{Q}}=0$ is precisely the Pfaff equation of the pfaffian form {\textit{heat}}, $\delta{\altmathcal{Q}}$ \cite{silvajunior}.}. Next, to seek more familiarity with the idea of integrable pfaffian forms, we will aim to analyze the solutions of the Pfaff equations associated with them. 

\begin{definition}\label{itm:1.4}{\normalfont(Pfaff Exact and Integrable Equation)}. The Pfaff equation associated with the pfaffian form $\delta{\xi}$, 
\begin{align*}
\delta{\xi}=0,
\end{align*}
is called exact if, and only if, $\delta{\xi}$ is an exact differential, $\delta{\xi}=d{\xi}$; if, and only if, the pfaffian form $\delta{\xi}$ is integrable, the associated Pfaff equation is called integrable.
\end{definition}

If $\delta{\xi}$ is a pfaffian form that constitutes an exact differential, then $\delta{\xi}=d{\xi}$ and the associated Pfaff equation $d{\xi}=0$ has as solution ${\xi}={\xi}(x_{1},x_{2},...,x_{n})$ = constant, which is geometrically a hypersurface of $n-1$ dimension in $B$.

On the other hand, if $\delta{\xi}$ is a pfaffian form that is an inexact differential but integrable, then $\delta{\xi}=0$ occurs in the same paths where $d\psi=0$, i.e., where $\delta{\xi}=\mu{d\psi}$ holds, according to Definition \ref{itm:1.2}. In this situation, the solution of $\delta{\xi}=0$ is $\psi=\psi(x_{1},x_{2},...,x_{n})$ = constant, which defines a hypersurface of $n-1$ dimension, now, in $A$. Of course, if $\delta{\xi}$ is a non-integrable Pfaffian form, then the solution of the equation $\delta{\xi}=0$ does not need to define any geometric object restricted to $n-1$ dimensions as in the previous cases. 

This said, we can now ask the main question: in which situations is a pfaffian form integrable? The sections \ref{sec:local} and \ref{sec:global} give us the answer.

\section{Local Integrability} \label{sec:local}

This section deals with integrability conditions that have \textit{local} character for pfaffian forms; i.e., conditions that, when satisfied, are restricted to some neighborhood $M$ contained in $B$, around some point $p$ of $B$. This will become clearer in the section \ref{sec:global}, where we will discuss conditions for \textit{global} integrability. For now, it is interesting that a definition for local integrability be formalized.

\begin{definition}\label{itm:2.1}{\normalfont(Local integrability)}. If the non-singular pfaffian form $\delta{\xi}$ is integrable restrictedly to some neighborhood $M\:{\subset}\:B$ of every point $p\:{\in}\:B$, we say that $\delta{\xi}$ is locally integrable on $B$.
\end{definition}

We will first discuss the simplest cases for local integrability: those of pfaffian forms in two and three variables.

\subsection{Pfaffian forms in two and three variables}

To avoid unnecessary repetition, we will henceforth assume only \textit{non-singular} pfaffian forms. According to Definition \ref{itm:1.2}, let be a pfaffian form $\delta{\xi}$ in two variables, ${x_1}$ and ${x_2}$:

\begin{equation}\label{eq:1}
\delta{\xi}=F_{1}({x_1},{x_2})dx_{1}+F_{2}({x_1},{x_2})dx_{2}.
\end{equation}

\noindent
The respective Pfaff equation associated with the pfaffian form of the expression (\ref{eq:1}) is

\begin{equation}\label{eq:2}
F_{1}({x_1},{x_2})dx_{1}+F_{2}({x_1},{x_2})dx_{2}=0,
\end{equation}

\noindent
which defines the following first-order ordinary differential equation,

\begin{equation}\label{eq:3}
\frac{d{x_2}}{d{x_1}}=-\frac{F_{1}({x_1},{x_2})}{F_{2}({x_1},{x_2})}\,{\equiv}\,f({x_1},{x_2}),
\end{equation}

\noindent
where $x_{2}=x_{2}(x_{1})$. Now, by the Existence and Uniqueness Theorem for ordinary differential equations\footnote{Attention spent on formally stating this theorem is redundant to the purpose of this paper. For more details of this fundamental theorem we suggest reading the book by Coddington and Levinson \cite{coddington}. }, if $f({x_1},{x_2})$ and ${\partial}f({x_1},{x_2})/{\partial}x_{2}$ are continuous on the open $B$, then given some point $p=({x_1}^{0},{x_2}^{0})\;{\in}B\;{\subseteq}\:{\mathbb {R}}^{2}$, there then exists in $B$ a single curve ${\psi}({x_1},x_{2}(x_{1}))=$ constant, parametrized by $x_1$, which provides the function ${x_2}=x_{2}({x_1})$ solution of the equation (\ref{eq:3}), such that it satisfies ${x_2}^{0}={x_2}({x_1}^{0})$ on some open interval $I$ containing ${x_1}^{0}$. We guarantee that the functions $F_{1}({x_1},{x_2})$ and $F_{2}({x_1},{x_2})$ are $C^{\infty}$ on $B$ by definition, and we must assume here that they are also, by construction, always non-null on $B$, given the arbitrariness generated by setting up the equation (\ref{eq:3}) so that $x_{2}=x_{2}(x_{1})$, instead of $x_{1}=x_{1}(x_{2})$. Otherwise, by this arbitrariness, it could be $\delta{\xi}$ identically null, or indeterminate. These collocations allow the use of this theorem for the equation (\ref{eq:3}). With this in mind, we are able to deal with one of the most important theorems in the theory of integrability of pfaffian forms.

\begin{theorem}\label{itm:1} Every pfaffian form in two variables on an open $B$ is locally integrable on $B$.
\end{theorem}

\noindent
\textbf{Proof.} Every Pfaff equation of a pfaffian form in two variables, whether this equation is exact or not, defines an first-order ordinary differential equation as the equation (\ref{eq:3}) that ensures, by the Existence and Uniqueness Theorem, at least \textit{locally}, the existence of a unique solution curve ${\psi}({x_1},x_{2}(x_{1}))=$ constant. Consider $d\psi$ in an open $B\:{\subseteq}\:{\mathbb {R}}^{2}$:

\begin{equation}\label{eq:4}
d\psi=\frac{{\partial}\psi}{{\partial}x_{1}}d{x_1}+\frac{{\partial}\psi}{{\partial}x_{2}}d{x_2}=0.
\end{equation}

Notice that equation (\ref{eq:4}) describes the same curves $x_{2}=x_{2}(x_{1})$ as equation (\ref{eq:3}). By then substituting equation (\ref{eq:3}) into equation (\ref{eq:4}), we have: 

\begin{equation}\label{eq:5}
d\psi=\frac{{\partial}\psi}{{\partial}x_{1}}d{x_1}-\frac{F_{1}({x_1},{x_2})}{F_{2}({x_1},{x_2})}\frac{{\partial}\psi}{{\partial}x_{2}}d{x_1}=0.
\end{equation}

Rearranging the equation (\ref{eq:5}), and in view of Definition \ref{itm:1.2}, we are invited to define the function ${\mu}={\mu}(x_{1},x_{2})$, clearly non-zero, given by,

\begin{equation}\label{eq:6}
{{\mu}(x_{1},x_{2})}^{-1}\,{\equiv}\,\frac{1}{F_{1}({x_1},{x_2})}\frac{{\partial}\psi}{{\partial}x_{1}}=\frac{1}{F_{2}({x_1},{x_2})}\frac{{\partial}\psi}{{\partial}x_{2}},
\end{equation}

\noindent
from which it immediately follows that:

\begin{equation}\label{eq:7}
{\mu}d{\psi}=F_{1}({x_1},{x_2})dx_{1}+F_{2}({x_1},{x_2})dx_{2}=\delta{\xi}.
\end{equation}
\hfill$\square$

Another proof for Theorem \ref{itm:1} can be found in \cite{diaz}. In studying the integrability of a pfaffian form $\delta{\xi}$ in three variables, $x_1$, $x_2$ and $x_3$,

\begin{equation}\label{eq:8}
\delta{\xi}=F_{1}({x_1},{x_2},{x_3})dx_{1}+F_{2}({x_1},{x_2},{x_3})dx_{2}+F_{3}({x_1},{x_2},{x_3})dx_{3},
\end{equation}

\noindent
it is more pertinent to use the vector notation: $\textbf{F}\,{\equiv}\,(F_{1},F_{2},F_{3})$, $d\textbf{r}\,{\equiv}\,(d{x_1},d{x_2},d{x_3})$. Thus, $\delta{\xi}$ and its associated Pfaff equation are represented by, respectively: $\delta{\xi}=\textbf{F}\,{\cdot}\,d\textbf{r}$ and $\textbf{F}\,{\cdot}\,d\textbf{r}=0$. In an open set, verification of the following equation is necessary and sufficient for the integrability of the pfaffian form in question:

\begin{equation}\label{eq:9}
\textbf{F}\,{\cdot}\,\nabla{\times}\textbf{F}=0.
\end{equation}

The preceding statement is a theorem. 	The proof of this result using the means discussed so far is long, in particular as to whether the equation (\ref{eq:9}) is sufficient for integrability, and so will be omitted in this paper. Later, when we deal with the integrability of pfaffian forms in any number of variables, the demonstration of the condition (\ref{eq:9}) for the case of three variables will be immediately retrieved. A fact to be pointed out now is that in the complete demonstration of the integrability condition (\ref{eq:9}) use is made of the Theorem \ref{itm:1} for the conclusion of the integrability of pfaffian forms in three variables \cite{sneddon}. As previously discussed, Theorem \ref{itm:1} has exclusively \textit{local} character. The result of this is that the condition (\ref{eq:9}) guarantees the integrability of pfaffian forms in three variables also only \textit{locally}, for an appropriate open $B\:{\subseteq}\:{\mathbb {R}}^{3}$.
\begin{theorem}\label{itm:2} A pfaffian form in three variables on an open $B$ is locally integrable on $B$ if, and only if, $\textbf{F}\;{\cdot}\,\nabla{\times}\textbf{F}=0$ on $B$.
\end{theorem}

\noindent
\textbf{Indication of proof.} Noting the Theorem \ref{itm:1}, see Chapter 1 of Sneddon's book \cite{sneddon}.
\hfill$\square$

However, it is not difficult to see that the equation (\ref{eq:9}) is a necessary condition for the integrability of a pfaffian form in three variables. Let us see, with the vector notation presented earlier, from the vector calculus we take that if $\nabla{\times}\textbf{F}\neq\textbf{0}$, then $\delta{\xi}$ is an inexact differential, because of Schwarz's Theorem. In order for $\delta{\xi}$ to be integrable, then there must exist a non-identically null function $\mu$ such that, $\nabla{\times}({\mu}^{-1}\textbf{F})=\textbf{0}$. In other words,

\begin{equation}\label{eq:10}
{\mu}^{-1}\nabla{\times}\textbf{F}+\nabla({\mu}^{-1}){\times}\textbf{F}=\textbf{0}.
\end{equation}

\noindent
By the scalar multiplication of (\ref{eq:10}) by $\textbf{F}$, we obtain that the condition sought is:

\begin{equation}\label{eq:11}
\textbf{F}\,{\cdot}\,\nabla{\times}\textbf{F}=0.
\end{equation}

For pfaffian forms in $n$ variables, we start by finding a necessary condition for integrability that generalizes (\ref{eq:11}). 

\subsection{Pfaffian forms in $n$ variables}

The important first results that follow were initially \cite{katz} obtained by Clairaut. 

\begin{lemma}\label{lemma1} A necessary condition for the integrability of a pfaffian form in n variables is that {} $\mathfrak{R}_{ijk}$:

\begin{align*}
\mathfrak{R}_{ijk}\:{\equiv}\:F_{i}\Bigg[\frac{{\partial}F_{k}}{{\partial}x_{j}}-\frac{{\partial}F_{j}}{{\partial}x_{k}}\Bigg]+F_{j}\Bigg[\frac{{\partial}F_{i}}{{\partial}x_{k}}-\frac{{\partial}F_{k}}{{\partial}x_{i}}\Bigg]+F_{k}\Bigg[\frac{{\partial}F_{j}}{{\partial}x_{i}}-\frac{{\partial}F_{i}}{{\partial}x_{j}}\Bigg],
\end{align*}

\noindent
be annulled, for any $i$, $j$, $k$.
\end{lemma}

\noindent
\textbf{Proof.} We start with writing a pfaffian form $\delta{\xi}$ in $n$ variables, $x_1, x_2,..., x_n$:

\begin{equation}\label{eq:12}
\delta{\xi}=\sum_{i=1}^{n}F_{i}(x_{1},x_{2},...,x_{n})dx_{i}.
\end{equation}

From Definition \ref{itm:1.2}, if $\delta{\xi}$ is integrable, then there exist functions $\mu$ and $\psi$ which, under the appropriate conditions, satisfy $\delta{\xi}=\mu{d{\psi}}$. It follows that, for each $i$:

\begin{equation}\label{eq:13}
\frac{{\partial}\psi}{{\partial}x_{i}}=\frac{1}{\mu}F_{i}.
\end{equation}

Now, if we derive the equation (\ref{eq:13}) with respect to some other variable, namely $x_j$, we will have,

\begin{equation}\label{eq:14}
\frac{{\partial}^{2}\psi}{{\partial}x_{j}{\partial}x_{i}}=\frac{{\partial}{\mu}^{-1}}{{\partial}x_{j}}F_{i}+{\mu}^{-1}\frac{{\partial}F_{i}}{{\partial}x_{j}}.
\end{equation}

\noindent
By Schwarz's Theorem, ${{\partial}^{2}\psi}/{{\partial}x_{j}{\partial}x_{i}}={{\partial}^{2}\psi}/{{\partial}x_{i}{\partial}x_{j}}$, so,

\begin{equation}\label{eq:15}
\frac{{\partial}{\mu}^{-1}}{{\partial}x_{i}}F_{j}+{\mu}^{-1}\frac{{\partial}F_{j}}{{\partial}x_{i}}=\frac{{\partial}{\mu}^{-1}}{{\partial}x_{j}}F_{i}+{\mu}^{-1}\frac{{\partial}F_{i}}{{\partial}x_{j}}
\end{equation}

\noindent
Regrouping (\ref{eq:15}) and multiplying the whole equation by $\mu$, 

\begin{equation}\label{eq:16}
\frac{{\partial}F_{j}}{{\partial}x_{i}}-\frac{{\partial}F_{i}}{{\partial}x_{j}}={\mu}F_{i}\frac{{\partial}{\mu}^{-1}}{{\partial}x_{j}}-{\mu}F_{j}\frac{{\partial}{\mu}^{-1}}{{\partial}x_{i}}.
\end{equation}

Compared to the condition (\ref{eq:11}) for three variables, the left-hand side of (\ref{eq:16}) invites us to look for ways to nullify it and thus obtain an integrability condition that depends only on the derivatives of the functions $F_i$. This occurs if we multiply (\ref{eq:16}) by another function $F_k$, and then cyclically add terms analogous to $F_k[{{\partial}F_{j}}/{{\partial}x_{i}}-{{\partial}F_{i}}/{{\partial}x_{j}}]$ so that,

\begin{equation}\label{eq:17}
\mathfrak{R}_{ijk}\,{\equiv}\,F_{i}\Bigg[\frac{{\partial}F_{k}}{{\partial}x_{j}}-\frac{{\partial}F_{j}}{{\partial}x_{k}}\Bigg]+F_{j}\Bigg[\frac{{\partial}F_{i}}{{\partial}x_{k}}-\frac{{\partial}F_{k}}{{\partial}x_{i}}\Bigg]+F_{k}\Bigg[\frac{{\partial}F_{j}}{{\partial}x_{i}}-\frac{{\partial}F_{i}}{{\partial}x_{j}}\Bigg]=0,
\end{equation}

\noindent
because the terms on the right-hand side of (\ref{eq:16}) cancel out with the analogous terms when we add them up. 
\hfill$\square$

Immediately one sees the recovery of condition (\ref{eq:11}) by setting $(i,j,k)=(1,2,3)$ on (\ref{eq:17}). The reciprocal of the Lemma \ref{lemma1} is valid, at least \textit{locally}. To show this, we first need to observe that: if the quantity $\mathfrak{R}_{ijk}$ is null in a collection of variables, it will remain null by a change of those variables.

\begin{lemma}\label{lemma2} The nullity of {} $\mathfrak{R}_{ijk}$ is invariant by a change of variables.
\end{lemma}

\noindent
\textbf{Proof.} Let be a pfaffian form $\delta{\xi}$ in $n$ variables $x_1, x_2,..., x_n$ such that it undergoes a change of variables to new $n$ variables ${\bar{x}_1}, {\bar{x}_2},..., {\bar{x}_n}$. The differential of a variable $x_{i}=x_{i}({\bar{x}_1}, {\bar{x}_2},..., {\bar{x}_n})$ is then:

\begin{equation}\label{eq:18}
d{x_i}=\sum_{j=1}^{n} \frac{{\partial}x_{i}}{{\partial}{\bar{x}}_{j}}d{\bar{x}}_{j}.
\end{equation}

The pfaffian form $\delta{\xi}$ can be represented in both collections of variables, with their associated functions. Hence,

\begin{equation}\label{eq:19}
\delta{\xi}=\sum_{i=1}^{n} {F_i}({x}_1, {x}_2,..., {x}_n)d{x}_{i}=\sum_{j=1}^{n} {\bar{F}}_{j}({\bar{x}_1}, {\bar{x}_2},..., {\bar{x}_n})d{\bar{x}}_{j},
\end{equation}

\noindent
where, substituting (\ref{eq:18}) into (\ref{eq:19}), we obtain:

\begin{equation}\label{eq:20}
{\bar{F}}_{j}({\bar{x}_1}, {\bar{x}_2},..., {\bar{x}_n})=\sum_{i=1}^{n}\frac{{\partial}x_{i}}{{\partial}{\bar{x}}_{j}}{F_i}({x}_1, {x}_2,..., {x}_n).
\end{equation}

Suppressing the explicit dependence to variables, by (\ref{eq:17}), in the collection of new variables, the quantity ${\bar{\mathfrak{R}}}_{ijk}$ is:

\begin{equation}\label{eq:21}
{\bar{F}}_{i}\Bigg[\frac{{\partial}{\bar{F}}_{k}}{{\partial}{\bar{x}}_{j}}-\frac{{\partial}{\bar{F}}_{j}}{{\partial}{\bar{x}}_{k}}\Bigg]+{\bar{F}}_{j}\Bigg[\frac{{\partial}{\bar{F}}_{i}}{{\partial}{\bar{x}}_{k}}-\frac{{\partial}{\bar{F}}_{k}}{{\partial}{\bar{x}}_{i}}\Bigg]+{\bar{F}}_{k}\Bigg[\frac{{\partial}{\bar{F}}_{j}}{{\partial}{\bar{x}}_{i}}-\frac{{\partial}{\bar{F}}_{i}}{{\partial}{\bar{x}}_{j}}\Bigg].
\end{equation}

We will initially analyze only the second term of (\ref{eq:21}), thus appropriately substituting the new functions given in (\ref{eq:20}). By doing so, 

\begin{equation}\label{eq:22}
\sum_{i=1}^{n}\frac{{\partial}x_{i}}{{\partial}{\bar{x}}_{j}}{F_i}\Bigg[\sum_{k=1}^{n}\sum_{j=1}^{n}\frac{{\partial}x_{k}}{{\partial}{\bar{x}}_{i}}\frac{{\partial}{F}_{k}}{{\partial}{x}_{j}}\frac{{\partial}x_{j}}{{\partial}{\bar{x}}_{k}}-\sum_{j=1}^{n}\sum_{k=1}^{n}\frac{{\partial}x_{j}}{{\partial}{\bar{x}}_{k}}\frac{{\partial}{F}_{j}}{{\partial}{x}_{k}}\frac{{\partial}x_{k}}{{\partial}{\bar{x}}_{i}}\Bigg],
\end{equation}

\noindent
we see that we obtain a term proportional to the first term of ${\mathfrak{R}}_{ijk}$, since the second order partial derivatives in the variables vanish, by Schwarz's Theorem.  Repeating the same for the remaining terms of (\ref{eq:21}), we have that:

\begin{equation}\label{eq:23}
{\bar{\mathfrak{R}}}_{ijk}=\Bigg[\sum_{i=1}^{n}\sum_{j=1}^{n}\sum_{k=1}^{n}\frac{{\partial}x_{i}}{{\partial}{\bar{x}}_{j}}\frac{{\partial}x_{j}}{{\partial}{\bar{x}}_{k}}\frac{{\partial}x_{k}}{{\partial}{\bar{x}}_{i}}\Bigg]{\mathfrak{R}}_{ijk}.
\end{equation}

\noindent
Therefore, if ${\mathfrak{R}}_{ijk}=0$, then ${\bar{\mathfrak{R}}}_{ijk}=0$.
\hfill$\square$

\begin{theorem}\label{itm:3} A sufficient condition for the local integrability of a pfaffian form in n variables, in an open B, is that {} $\mathfrak{R}_{ijk}$ annuls, for any $i$, $j$, $k$.
\end{theorem}

\noindent
\textbf{Proof.} We will present a proof via finite induction which initially seeks to show that the condition $\mathfrak{R}_{ijk}=0$, for any $i$, $j$, $k$, is sufficient for integrability. From this it will follow that the most we can say about such a condition is that, in fact, it is sufficient to \textit{local} integrability, only.

For a pfaffian form in one variable, $x_1$, by construction, it is clear that $\delta{\xi}$ is always integrable. Whereas for a pfaffian form in $n$ variables, 

\begin{equation}\label{eq:24}
\delta{\xi}=\sum_{i=1}^{n}F_{i}(x_{1},x_{2},...,x_{n})dx_{i},
\end{equation}

\noindent
we assume that ${\mathfrak{R}}_{ijk}=0$, for any $i,j,k$. Next, we choose to examine $\delta{\xi}$ for a path in an open $B\;{\subseteq}\;{\mathbb{R}}^{n}$ such that $d{x_n}=0$. The pfaffian form that results from fixing $x_n$ at (\ref{eq:24}), is,

\begin{equation}\label{eq:25}
{\delta}\eta=\sum_{i=1}^{n-1}F_{i}(x_{1},x_{2},...,x_{n})dx_{i},
\end{equation}

\noindent
so that naturally ${\mathfrak{R}}_{ijk}=0$ remains unchanged in (\ref{eq:25}), as the induction hypothesis, since the nullity of this quantity does not change because we fix a variable. We then assume that ${\delta}\eta$ is integrable, under the circumstance ${\mathfrak{R}}_{ijk}=0$, for any $i,j,k$ different from $n$ in $B$. On account of this, there must exist functions $\lambda$, with $\lambda(x_{1},x_{2},...,x_{n-1})\;{\neq}\;0$, and ${\sigma}={\sigma}(x_{1},x_{2},...,x_{n-1})$, in some open $A\:{\subseteq}\:B$, such that:

\begin{equation}\label{eq:26}
{\delta}\eta=\lambda{d\sigma}=\lambda\sum_{i=1}^{n-1}\frac{{\partial}\sigma}{{\partial}{x_i}}d{x_i}.
\end{equation}

Now, by letting $x_n$ vary, we can rewrite ${\delta}\xi$ as a function of ${\delta}\eta$, with ${\delta}\eta$ integrable by hypothesis, as we put it. That is,

\begin{equation}\label{eq:27}
{\delta}\xi=\lambda{d\sigma}+{F_n}d{x_n},
\end{equation}

\noindent
where, since ${\delta}\xi$ is a pfaffian form in $n$ variables, writing (\ref{eq:27}) is equivalent to a change of variables in ${\delta}\xi$, from the variables $x_1, x_2,..., x_n$, to certain new variables ${\bar{x}_1}, {\bar{x}_2},..., {\bar{x}_{n-2}},{\sigma},{x_n}$. In this new collection of variables it occurs that ${{\bar{F}}_i}=0$, for all $i=\{1,2,...,n-2\}$. From the Lemma \ref{lemma2}, the hypothesis of the nullity of ${\mathfrak{R}}_{ijk}$ holds for the new collection of variables. Again, this relationship is preserved when examining only the collection of variables ${\bar{x}_1}, {\bar{x}_2},..., {\bar{x}_{n-2}}$. Explicitly,

\begin{equation}\label{eq:28}
{\bar{\mathfrak{R}}}_{ijk}={\lambda}\frac{{\partial}{F_n}}{{\partial}{\bar{x}}_{i}}-{F_n}\frac{{\partial}\lambda}{{\partial}{\bar{x}}_{i}}=0,
\end{equation}

\noindent
and we obtain that, on the variables ${\bar{x}_1}, {\bar{x}_2},..., {\bar{x}_{n-2}},{\sigma},{x_n}$, the quotient ${F_n}/\lambda$ must depend solely on $\sigma$ and $x_n$. With $\lambda\;{\neq}\;0$, we can rewrite (\ref{eq:27}) as:

\begin{equation}\label{eq:29}
{\delta}\xi=\lambda\Bigg({d\sigma}+\frac{F_n}{\lambda}d{x_n}\Bigg),
\end{equation}

The term in parentheses in (\ref{eq:29}) is a pfaffian form in two variables, which is, by Theorem \ref{itm:1}, locally integrable. Therefore, there exist functions $\mu$ and $\psi$ such that,

\begin{equation}\label{eq:30}
{\delta}\xi=\lambda\mu{d\psi},
\end{equation}

\noindent
under the appropriate conditions, and so ${\delta}\xi$ is locally integrable on $B$.
\hfill$\square$

We will now present one last result. Originally obtained in the formalization of Classical Thermodynamics by C. Carathéodory in 1909 \cite{caratheodory}, and probably figuring as the criterion for integrability of pfaffian forms most absent from differential equations textbooks since then. This is a verification that provides the local integrability of a pfaffian form from a topological condition of the set $B$ to which the pfaffian form in question resides. The proof of this Carathéodory's Theorem is presented here in a little more detail than in the works that first investigated it \cite{buchdahl, buchdahl2}, after Carathéodory's original proof \cite{caratheodory}. 

\begin{theorem}\label{itm:4} {\normalfont(Carathéodory's Theorem)} A necessary and sufficient condition for the local integrability of a pfaffian form ${\delta}{\xi}$ in n variables in an open B, is that in every neighborhood $M\:{\subset}{\:}B$ arbitrarily close to any point $p\:{\in}{\:}B$ there exist points unreachable from $p$ by curves such that ${\delta}{\xi}=0$.
\end{theorem}

\noindent
\textbf{Proof.} \cite{buchdahl2} Let be the pfaffian form ${\delta}\xi$ in $n$ variables, in an open $B\:{\subseteq}\:{\mathbb {R}}^{n}$, and $p=({x_1}^{0},{x_2}^{0},...,{x_n}^{0})$, $q=({x_1}^{*},{x_2}^{*},...,{x_n}^{*})$, $r=({x_1}^{**},{x_2}^{**},...,{x_n}^{**})$ points of $B$. Let be the curves ${\gamma}_1$ and ${\gamma}_2$ in $B$, smooth, parametrized by a real parameter $t$, 

\begin{subequations}\label{31} 
\begin{gather}
{\gamma}_{1}(t)=(f_{1}(t),f_{2}(t),...,f_{n}(t))\,{\equiv}{\,}(f_{i}(t)), \label{31a}\\
{\gamma}_{2}(t)=(f_{1}(t)+{\nu}g_{1}(t),f_{2}(t)+{\nu}g_{2}(t),...,f_{n}(t)+{\nu}g_{n}(t))\,{\equiv}{\,}(f_{i}(t)+{\nu}g_{i}(t)), \label{31b}
\end{gather}
\end{subequations}

\noindent
with ${\nu}$ real, such that ${\delta}{\xi}=0$, with the following conditions, respectively,

\begin{subequations} \label{32}
\begin{align}
{\gamma}_{1}(t_{0})=&{\,}p,\quad {\gamma}_{1}(t_{*})=q, \label{32a}\\
{\gamma}_{2}(t_{0})=&{\,}p,\quad {\gamma}_{2}(t_{**})=r, \label{32b}
\end{align}
\end{subequations}

\noindent
where $t_{0}<t_{*}<t_{**}$, with $|t_{*}-t_{0}|<\epsilon_{1}$ and $|t_{**}-t_{*}|<\epsilon_{2}$, for arbitrarily small $\epsilon_{1}$ and $\epsilon_{2}$. With a sufficiently small $\nu$, our goal is to examine the situation $\epsilon_{2}\;{\rightarrow}\;0$. The Pfaff equation to which ${\gamma}_{2}$ is solution is given by,

\begin{equation} \label{33}
\begin{aligned}
&\sum_{i=1}^{n}F_{i}(f_{l}(t)+{\nu}g_{l}(t))d(f_{i}(t)+{\nu}g_{i}(t))=0\\
=&\sum_{i=1}^{n}F_{i}(f_{l}(t)+{\nu}g_{l}(t))[\dot{f}_{i}(t)+{\nu}\dot{g}_{i}(t)],
\end{aligned}
\end{equation}

\noindent
with $d{f_i}(t)/dt\,{\equiv}{\,}\dot{f}_{i}(t)$, $d{g_i}(t)/dt\,{\equiv}{\,}\dot{g}_{i}(t)$. Deriving (\ref{33}) by $\nu$, at $\nu=0$,

\begin{equation}\label{eq:34}
\sum_{i=1}^{n}F_{i}(f_{l}(t))\dot{g}_{i}(t)+\sum_{i=1}^{n}\sum_{j=1}^{n}\frac{{\partial}F_{i}(f_{l}(t))}{{\partial}{x_j}}\dot{f}_{i}(t){g}_{j}(t)=0,
\end{equation}

\noindent
or

\begin{equation}\label{eq:35}
\sum_{i=1}^{n}F_{i}(f_{l}(t))\dot{g}_{i}(t)=-\sum_{i=1}^{n}\sum_{j=1}^{n}\frac{{\partial}F_{i}(f_{l}(t))}{{\partial}{x_j}}\dot{f}_{i}(t){g}_{j}(t).
\end{equation}

Equation (\ref{eq:35}) is equivalent to us choosing $n-1$ of the functions ${g}_{j}(t)$ in an arbitrary manner and the $n$-th one we ensure obeys (\ref{eq:35}). Let then be that $n$-th function ${g}_{k}(t)$, so that, isolating the $k$-index terms, we have:

\begin{equation}\label{eq:36}
\begin{aligned}
F_{k}(f_{l}(t))\dot{g}_{k}(t)+&\sum_{\substack{i=1 \\ i\neq k}}^{n}\frac{{\partial}F_{i}(f_{l}(t))}{{\partial}{x_k}}\dot{f}_{i}(t){g}_{k}(t)=\\-&\sum_{\substack{j=1 \\ j\neq k}}^{n}F_{j}(f_{l}(t))\dot{g}_{j}(t)-\sum_{\substack{i=1 \\ i\neq k}}^{n}\sum_{\substack{j=1 \\ j\neq k}}^{n}\frac{{\partial}F_{i}(f_{l}(t))}{{\partial}{x_j}}\dot{f}_{i}(t){g}_{j}(t).
\end{aligned}
\end{equation}

The function ${g}_{k}(t)$ then becomes the object to be studied if we make $\epsilon_{2}\;{\rightarrow}\;0$. The left-hand side of (\ref{eq:36}) is a first-order linear ordinary differential equation in ${g}_{k}(t)$, so, by Leibniz's method of the integrating factor, let be the function ${\eta}={\eta}(t)$, non-null, such that:

\begin{equation}\label{eq:37}
\frac{d({\eta}(t)F_{k}(f_{l}(t)){g}_{k}(t))}{dt}={\eta}(t)\Bigg[F_{k}(f_{l}(t))\dot{g}_{k}(t)+\sum_{\substack{i=1 \\ i\neq k}}^{n}\frac{{\partial}F_{i}(f_{l}(t))}{{\partial}{x_k}}\dot{f}_{i}(t){g}_{k}(t)\Bigg].
\end{equation}

\noindent
Developing the left-hand side of (\ref{eq:37}),

\begin{equation}\label{eq:38}
\begin{aligned}
\frac{d({\eta}(t)F_{k}(f_{l}(t)){g}_{k}(t))}{dt}&\\={\eta}(t)F_{k}(f_{l}(t))\dot{g}_{k}(t)+&{\eta}(t)\sum_{\substack{i=1 \\ i\neq k}}^{n}\frac{{\partial}F_{k}(f_{l}(t))}{{\partial}{x_i}}\dot{f}_{i}(t){g}_{k}(t)+\dot{{\eta}}(t)F_{k}(f_{l}(t)){g}_{k}(t),
\end{aligned}
\end{equation}

\noindent
and with the comparison between (\ref{eq:37}) and (\ref{eq:38}) comes,

\begin{equation}\label{eq:39}
\dot{{\eta}}(t)F_{k}(f_{l}(t))={\eta}(t)\dot{f}_{i}(t)\sum_{\substack{i=1 \\ i\neq k}}^{n}\Bigg[\frac{{\partial}F_{i}(f_{l}(t))}{{\partial}{x_k}}-\frac{{\partial}F_{k}(f_{l}(t))}{{\partial}{x_i}}\Bigg].
\end{equation}

Now, perpetuating Leibniz's method and multiplying the two sides of (\ref{eq:36}) by ${\eta}(t)$, using (\ref{eq:38}) and (\ref{eq:39}), we can isolate ${g}_{k}(t)$ by integrating the left-hand side of (\ref{eq:37}), which turns out to be equal to the right-hand side of (\ref{eq:36}) when we multiply the latter by ${\eta}(t)$. We then have:

\begin{equation}\label{eq:40}
{\eta}(t')F_{k}(f_{l}(t')){g}_{k}(t')=-\int_{t_0}^{t'}{\eta}(t)\Bigg[\sum_{\substack{j=1 \\ j\neq k}}^{n}F_{j}(f_{l}(t))\dot{g}_{j}(t)+\sum_{\substack{j=1 \\ j\neq k}}^{n}\sum_{\substack{i=1 \\ i\neq k}}^{n}\frac{{\partial}F_{i}(f_{l}(t))}{{\partial}{x_j}}\dot{f}_{i}(t){g}_{j}(t)\Bigg]dt,
\end{equation}

\noindent
where $g_i(t_0)=0$ for all $i$, by the conditions (\ref{32}). Integrating by parts the first term in the integrand of (\ref{eq:40}), using (\ref{eq:39}) and again that $g_i(t_0)=0$, we obtain directly:

\begin{equation}\label{eq:41}
\begin{aligned}
&-\int_{t_0}^{t'}{\eta}(t)\sum_{\substack{j=1 \\ j\neq k}}^{n}F_{j}(f_{l}(t))\dot{g}_{j}(t)dt=-{\eta}(t')\sum_{\substack{j=1 \\ j\neq k}}^{n}F_{j}(f_{l}(t')){g}_{j}(t')\\&+\int_{t_0}^{t'}{\eta}(t)\sum_{\substack{i=1 \\ i\neq k}}^{n}\sum_{\substack{j=1 \\ j\neq k}}^{n}\dot{f}_{i}(t){g}_{j}(t)\Bigg(\frac{F_{j}(f_{l}(t))}{F_{k}(f_{l}(t))}\Bigg[\frac{{\partial}F_{i}(f_{l}(t))}{{\partial}{x_k}}-\frac{{\partial}F_{k}(f_{l}(t))}{{\partial}{x_i}}\Bigg]+\frac{{\partial}F_{j}(f_{l}(t))}{{\partial}{x_i}}\Bigg)dt.
\end{aligned}
\end{equation}

\noindent
Substituting (\ref{eq:41}) into (\ref{eq:40}), and putting the function $F_{k}(f_{l}(t))$ in evidence on the integrand, we have:

\begin{equation}\label{eq:42}
\begin{aligned}
{\eta}(t')F_{k}(f_{l}(t')){g}_{k}(t')=&-{\eta}(t')\sum_{\substack{j=1 \\ j\neq k}}^{n}F_{j}(f_{l}(t')){g}_{j}(t')\\&+\sum_{\substack{i=1 \\ i\neq k}}^{n}\sum_{\substack{j=1 \\ j\neq k}}^{n}\int_{t_0}^{t'}\frac{{\eta}(t)\dot{f}_{i}(t){g}_{j}(t)}{F_{k}(f_{l}(t))}\Bigg(F_{j}(f_{l}(t))\Bigg[\frac{{\partial}F_{i}(f_{l}(t))}{{\partial}{x_k}}-\frac{{\partial}F_{k}(f_{l}(t))}{{\partial}{x_i}}\Bigg]\\&+F_{k}(f_{l}(t))\Bigg[\frac{{\partial}F_{j}(f_{l}(t))}{{\partial}{x_i}}-\frac{{\partial}F_{i}(f_{l}(t))}{{\partial}{x_j}}\Bigg]\Bigg)dt.
\end{aligned}
\end{equation}

\noindent
Since, by the curve $\gamma_{1}$ it is true that,

\begin{equation}\label{eq:43}
\sum_{i=1}^{n}F_{i}(f_{l}(t))\dot{f}_{i}(t)=0,
\end{equation}

\noindent
then,

\begin{equation}\label{eq:44}
\sum_{\substack{i=1 \\ i\neq k}}^{n}\sum_{\substack{j=1 \\ j\neq k}}^{n}\int_{t_0}^{t'}\frac{{\eta}(t)\dot{f}_{i}(t){g}_{j}(t)}{F_{k}(f_{l}(t))}F_{i}(f_{l}(t))\Bigg[\frac{{\partial}F_{k}(f_{l}(t))}{{\partial}{x_j}}-\frac{{\partial}F_{j}(f_{l}(t))}{{\partial}{x_k}}\Bigg]dt=0.
\end{equation}

Substituting (\ref{eq:44}) into (\ref{eq:42}), isolating ${g}_{k}(t')$ in the process, and identifying ${\mathfrak{R}}_{ijk}$ in the integrand, we have: 

\begin{equation}\label{eq:45}
\begin{aligned}
{g}_{k}(t')=\frac{1}{{\eta}(t')F_{k}(f_{l}(t'))}\Bigg\{&-{\eta}(t')\sum_{\substack{j=1 \\ j\neq k}}^{n}F_{j}(f_{l}(t')){g}_{j}(t')\\&+\sum_{\substack{i=1 \\ i\neq k}}^{n}\sum_{\substack{j=1 \\ j\neq k}}^{n}\int_{t_0}^{t'}\frac{{\eta}(t)\dot{f}_{i}(t){g}_{j}(t)}{F_{k}(f_{l}(t))}{\mathfrak{R}}_{ijk}dt\Bigg\}.
\end{aligned}
\end{equation}

If $t'{\rightarrow}{t_{*}}$, with $\epsilon_{1}$ arbitrarily small, where $|t_{*}-t_{0}|<\epsilon_{1}$, as well as $\epsilon_{2}$, where $|t_{**}-t_{*}|<\epsilon_{2}$, then, in the limit $\epsilon_{2}\;{\rightarrow}{\;}0$, we will have the formation of a neighborhood $M{\:}{\subset}{\:}B$ around $p$, arbitrarily close to $p$, such that there are points reachable from $p$ by curves on which ${\delta}{\xi}=0$. The only exception to this is if every function ${g}_{k}(t)$ is identically null, for all $t$. In the integrand of (\ref{eq:45}), this requires that,

\begin{equation}\label{eq:46}
\sum_{\substack{i=1 \\ i\neq k}}^{n}\dot{f}_{i}(t){\mathfrak{R}}_{ijk}=0,
\end{equation}

\noindent
since, by construction, ${\eta}={\eta}(t){\:}{\neq}{\:}0$, the ${g_j}(t)$ cannot be fixed as zero and, obviously, ${F_{k}(f_{l}(t))}^{-1}{\:}{\neq}{\:}0$, for all $k$. Since the $\dot{f}_{i}(t)$ can be arbitrary, except $\dot{f}_{k}(t)$, which does not appear in (\ref{eq:46}), we conclude that ${\mathfrak{R}}_{ijk}=0$, for any $i$, $j$, $k$. By the Lemma \ref{lemma1} and the Theorem \ref{itm:3} this proof ends.
\hfill$\square$

\section{Global Integrability}\label{sec:global}
The Carathéodory's Theorem \ref{itm:4} has historically been placed under debate for guaranteeing only a \textit{local} integrability for pfaffian forms \cite{boyling}. However, its use in Classical Thermodynamics provides clues that the local nature of the Carathéodory's Theorem lies in the generality of its premise in topological terms. More than that, when analyzed according to the descriptive needs of Classical Thermodynamics \cite{silvajunior}, this theorem appears to ask for more than it needs to obtain an integrating factor, by presuming a non-connection relation between points in space (in the present context, the ${\mathbb {R}}^{n}$) valid for any, arbitrarily small, neighborhood in this space.  

This indicates the possibility of a hidden content in Carathéodory's Theorem that can be revealed by appropriate modification of the notion of \textit{neighborhood}. We will do this next, and obtain as a result a sufficient condition for the integrability of pfaffian forms on all $B$, except for a set of measure zero contained in $B$; what we will call here \textit{global} integrability on $B$. 

\begin{definition}\label{itm:3.1}{\normalfont(Surrounding line)}. Given a point $p=({x_1}^{0},{x_2}^{0},...,{x_n}^{0})\:{\in}{\:}B\:{\subseteq}{\:}{\mathbb {R}}^{n}$, for the line ${\Pi}(x_{i})$ of the points $({x_1}^{0},{x_2}^{0},...,{x_i},...,{x_n}^{0})$, with the arbitrary variable $x_i$ called the free variable, we call the surrounding line to $p$ associated with $x_i$.
\end{definition}

Notice that the union $\bigcup_{i=1}^{n} {\Pi}(x_{i})$ of the $n$ possible surrounding lines ${\Pi}(x_{i})$ to a point $p\:{\in}{\:}B\:{\subseteq}{\:}{\mathbb {R}}^{n}$ does not constitute neighborhood  $M$ of $p$.

\begin{theorem}\label{itm:5} A sufficient condition for the global integrability of a pfaffian form ${\delta}\xi$ in n variables in an open B is that on the surrounding line {} ${\Pi}(x_{i})$ of any point $p\:{\in}{\:}B$, for some free variable $x_{i}$, there exist points arbitrarily close to $p$ unreachable from $p$ by curves such that ${\delta}\xi=0$.
\end{theorem}

\noindent
\textbf{Proof.} Let be the pfaffian form ${\delta}\xi$ in $n$ variables, on an open $B\:{\subseteq}\:{\mathbb {R}}^{n}$, and $p=({x_1}^{0},{x_2}^{0},...,{x_n}^{0})$ an arbitrary point of $B$. Let $q=({x_1}^{0},{x_2}^{0},...,{x_n}^{*})$ also be a point on the surrounding line to $p$ associated with the variable $x_n$, ${\Pi}(x_{n})$, and ${\gamma}$ a curve in $B$ such that:

\begin{equation}\label{eq:47}
\sum_{i=1}^{n}{F_i}({\gamma})d{x_i}({\gamma})=0.
\end{equation}

With $|{x_n}^{*}-{x_n}^{0}|<{\varepsilon}$, let us assume that $\gamma$ passes through $p$ but does not pass through $q$, for any arbitrarily small $\varepsilon$. It follows from this that the equation,

\begin{equation}\label{eq:48}
d{x_n}({x_1},{x_2},...,{x_{n-1}})=-\sum_{i=1}^{n-1}\frac{{F_i}({x_1},{x_2},...,{x_n})}{{F_n}({x_1},{x_2},...,{x_n})}d{x_i},
\end{equation}

\noindent
arising from (\ref{eq:47}), denotes $d{x_n}$ as the differential of a function ${x_n}={x_n}({x_1},{x_2},...,{x_{n-1}})$, so since, as we know, the function $F_n({x_1},{x_2},...,{x_n})$ is not identically null. We obtain ${x_n}={x_n}({x_1},{x_2},...,{x_{n-1}})$ explicitly by integrating (\ref{eq:48}),

\begin{equation}\label{eq:49}
{x_n}({x_1},{x_2},...,{x_{n-1}})={x_n}^{0}-\int_{({x_1}^{0},{x_2}^{0},...,{x_{n-1}}^{0})}^{({x_1},{x_2},...,{x_{n-1}})}\sum_{i=1}^{n-1}\frac{{F_i}({x_1},{x_2},...,{x_n})}{{F_n}({x_1},{x_2},...,{x_n})}d{x_i},
\end{equation}

\noindent
where ${x_n}^{0}={x_n}({x_1}^{0},{x_2}^{0},...,{x_{n-1}}^{0})$. Now, if we make the quantities ${x_1}^{0},{x_2}^{0},...,{x_{n-1}}^{0}$ vary, we get ${x_1},{x_2},...,{x_n}^{0}$ as the new independent variables on which the function ${x_n}$, now ${x_n}={x_n}({x_1},{x_2},...,{x_n}^{0})$, depends. The function ${x_n}$ is continuous with respect to the variables ${x_1},{x_2},...,{x_{n-1}},{x_n}^{0}$, and differentiable with respect to the variables ${x_1},{x_2},...,{x_{n-1}}$, by (\ref{eq:48}). Hence:

\begin{equation}\label{eq:50}
\frac{{\partial}{x_n}}{{\partial}{x_i}}=-\frac{F_i}{F_n}.
\end{equation}

Furthermore, by equation (\ref{eq:49}), the quotients ${F_i}/{F_n}$ might depend on ${x_n}^{0}$ in some way. However, if we fix the other variables ${x_1},{x_2},...,{x_{n-1}}$, at (\ref{eq:49}), we get that ${x_n}={x_n}({x_1},{x_2},...,{x_n}^{0})$ is a monotone function of ${x_n}^{0}$. This does not change for any closed interval contained in ${\Pi}(x_{n})$ on which we can make the same assumptions that have been posited so far. As a consequence, by Lebesgue's Differentiation Theorem \cite{botsko}, ${x_n}={x_n}({x_1},{x_2},...,{x_n}^{0})$ is a differentiable function of ${x_n}^{0}$ on all $B$, except for a set of measure zero contained in $B$. Thus, we observe that the differential of the function ${x_n}={x_n}({x_1},{x_2},...,{x_n}^{0})$,

\begin{equation}\label{eq:51}
d{x_n}({x_1},{x_2},...,{x_n}^{0})=\sum_{i=1}^{n-1}\frac{{\partial}{x_n}}{{\partial}{x_i}}d{x_i}+\frac{{\partial}{x_n}}{{\partial}{x_n}^{0}}d{x_n}^{0},
\end{equation}

\noindent
in the same way refers to all $B$, except for the same set of measure zero contained in $B$. Retaking (\ref{eq:47}) explicitly, and using (\ref{eq:51}) and (\ref{eq:50}), we have:

\begin{equation}\label{eq:52}
\begin{aligned}
{\delta}{\xi}=&\sum_{i=1}^{n-1}{F_i}d{x_i}+{F_n}d{x_n}\\=&\sum_{i=1}^{n-1}{F_i}d{x_i}+{F_n}\Bigg(\sum_{i=1}^{n-1}\frac{{\partial}{x_n}}{{\partial}{x_i}}d{x_i}+\frac{{\partial}{x_n}}{{\partial}{x_n}^{0}}d{x_n}^{0}\Bigg)\\=&{F_n}\frac{{\partial}{x_n}}{{\partial}{x_n}^{0}}d{x_n}^{0}.
\end{aligned}
\end{equation}

\noindent
Therefore, ${\delta}{\xi}$ is integrable over almost everywhere on $B$.
\hfill$\square$

\section{Applications and concluding remarks}\label{sec:finais}

In this paper we introduce the reader to the integrability conditions of pfaffian forms on ${\mathbb {R}}^{n}$, with except for the well-known Frobenius Theorem. We divide our discussion between local aspects, in section \ref{sec:local}, and global aspects, in section \ref{sec:global}, with respect to integrability. Inspired by Carathéodory's Theorem, and its use in Classical Thermodynamics \cite{silvajunior}, an integrability criterion of \textit{global} character, namely, the Theorem \ref{itm:5}, was obtained in section \ref{sec:global}.

The Theorem \ref{itm:5} when applied on Classical Thermodynamics may generate a very important result: the construction of a differentiable almost everywhere entropy function, so that the regions which the differentiability vanish are, by the experimental justification of the theory, the points on thermodynamic space of states related to phase transitions. Indeed, introducing the zeroth law of thermodynamics and consequently the concept of a empirical temperature $\vartheta$, this quantity can be now identified how the free variable in the context of Theorem \ref{itm:5}; i.e., we assume that $\vartheta=x_n$. Next, in according with \cite{buchdahl2}, by identifying also ${\delta}{\xi}$ as the quantity of heat ${\delta}{\altmathcal{Q}}$, ${F_n}={F_{\vartheta}}$, and ${x_n}^{0}={\vartheta}^{0}$, the last expression in (\ref{eq:52}) can be rewritten if one observes the deduction of the premise of Theorem \ref{itm:5} from Kelvin's, or Clausius's, statement of the second law of thermodynamics \cite{silvajunior}. Then taking a change for new variables, ${\mu}$ and $\sigma$, one obtains, 

\begin{equation*}
{\delta}{\altmathcal{Q}}={\mu}d{\sigma},
\end{equation*}

\noindent
where it is not difficult to check that $\sigma$ is an differentiable almost everywhere function of appropriate variables under consideration. With a little more thermodynamics arguments \cite{buchdahl2}, one can conclude that  

\begin{equation*}
{\delta}{\altmathcal{Q}}=Td{\altmathcal{S}},
\end{equation*}

\noindent
being $T$ the absolute temperature, and ${\altmathcal{S}}$ the absolute entropy. This absolute entropy function, or just entropy function, have the following properties: a) additivity (over thermodynamics systems); b) extremization (maximization or minimization\footnote{For the reader with familiarity in physics, this possibility of maximization or minimization, in addition with the absence of a property of monotone increasing variation with respect the internal energy, actually are the blessings of Carathéodory's thermodynamics; there is a wide set of experimental evidences that supports this apparent gaps in the theory \cite{lavis}.}) on a irreversible adiabatic process (the premise of Theorem \ref{itm:5} can be generalized from the thermodynamic point of view of irreversible processes); c) differentiability almost everywhere. With the experimental based assumption of local bounded variation of the thermodynamics quantities \cite{lieb}, we can assume that the entropy function have also its first-derivatives with the local bounded variation property. This, together with the property of differentiability almost everywhere, generates a final property for the entropy function in question: d) local Lipschitz continuity. Actually, due to Rademacher's Theorem, the properties of this entropy function can be summarized in: additivity, extremization, and local Lipschitz continuity.  

In Analytical Mechanics as well, the application of a integrability criterion for pfaffian forms is the procedure which leads to the verification of whether non-holonomic constraints are, or are not, integrable. Constraints are relations between the mechanical generalized coordinates, generalized velocities, and eventually the time coordinate; when a constraint is a relation between only the generalized coordinates and time, it is called holonomic, otherwise it is called non-holonomic. A set of $n$ non-holonomic constraints imposed to a mechanical system are often represented by $n$ correspondents Pfaff equations: $\delta{{\xi}_{1}}=0,...,\delta{{\xi}_{n}}=0$. There are a practical importance for Analytical Mechanics in the verification of the integrability of non-holonomic constrains: the constraints can be applied directly in the lagrangian function of the mechanical system, thus making it easier to obtain the equations of motion by solving the Euler-Lagrange equations \cite{nivaldo}. Besides that, the traditional method to do the integrability test is by using the Frobenius Theorem, which calls for handling with exterior algebra and the analytical knowledge of the Pfaff equations in question.

On the other hand, in terms of original Carathéodory's Theorem the test of integrability for non-holonomic constraints can be performed without more mathematical features, instead making use of physical inferences in the phase space of the mechanical system and its respective constraints. For example, the traditional problem of a perfect cylinder rolling without sliding on an inclined plane can be easily visualized from the perspective that there are states in the phase space of the system that are not accessible, hence the respective constraint must be integrable. In this point of view, the original Carathéodory's Theorem show itself to be more physically substantial than Frobenius Theorem, although the latter have a greater rigour and be most useful, specially with more complicated constraints. However, the application and respective descriptive consequences of a specie of Carathéodory's Theorem, like Theorem \ref{itm:5}, in the integrability of non-holonomic constraint must be better investigate.


\begin{thebibliography}{18}

\bibitem{flanders} Flanders, H. (1989). \textit{Differential Forms with Applications to the Physical Sciences}. New York, NY: Dover Publications.

\bibitem{morita} Morita, S. (2001). \textit{Geometry of Differential Forms}. Providence, RI: American Mathematical Society.

\bibitem{antoniou} Antoniou, I. (2002). Caratheodory and the Foundations of Thermodynamics and Statistical Physics. \textit{Foundations of Physics}. 32(4): 627–641. doi.org/10.1023/A:1015040501205

\bibitem{arens} Arens, R. (1964). Differential-Geometric Elements of Analytic Dynamics. \textit{Journal of Mathematical Analysis
and Applications}. 9: 165–202.

\bibitem{katz} Katz, V. J. (1981). The history of differential forms from Clairaut to Poincaré. \textit{Historia Mathematica}. 8(2): 161–188.

\bibitem{samelson} Samelson, H. (2001). Differential Forms, the Early Days; or the Stories of Deahna’s Theorem and of Volterra’s Theorem. \textit{The American Mathematical Monthly}. 108(6): 522–530. doi.org/10.1080/00029890.2001.11919779

\bibitem{frobenius} Frobenius, G. (1877). Über das Pfaffsche Problem. \textit{Journal für die reine und angewandte Mathematik}. 82: 230–315.

\bibitem{cartan} Cartan, E. (1899). Sur certaines expressions différentielles et le problème de Pfaff. \textit{Annales scientifiques de
l’École Normale Supérieure}. 16: 239–332.

\bibitem{papastavridis} Papastavridis, J. G. (2002). \textit{Analytical Mechanics: A Comprehensive Treatise on the Dynamics of Constrained
Systems; For Engineers, Physicists, and Mathematicians}. Oxford, Oxon: Oxford University Press.

\bibitem{silvajunior} Silva Júnior, P. F. (2021). Sobre a Dedução do Axioma de Carathéodory da Segunda Lei da Termodinâmica dos Princípios de Clausius e Kelvin. \textit{Revista Brasileira de Ensino de Física}. 43. doi.org/10.1590/1806-9126-RBEF-2020-0448

\bibitem{coddington} Coddington, E. A., Levinson, N. (1955). \textit{Theory of Ordinary Differential Equations}. New York, NY: McGraw-Hill Inc.

\bibitem{diaz} Díaz, A. A. (2017). Las ecuaciones de Pfaff. Monograph. Universidad de Sevilla, Sevilla, ES.

\bibitem{sneddon} Sneddon, I. N. (2006). \textit{Elements of Partial Differential Equations}. New York, NY: Dover Publications.

\bibitem{caratheodory} Carathéodory, C. (1909). Untersuchungen über die Grundlagen der Thermodynamik. \textit{Mathematische Annalen}. 67: 355–386. doi.org/10.1007/BF01450409

\bibitem{buchdahl} Buchdahl, H. A. (1954). Integrability Conditions and Carathéodory’s Theorem. \textit{American Journal of Physics}. 22: 182–183. doi.org/10.1119/1.1933675

\bibitem{buchdahl2} Buchdahl, H. A. (1966). \textit{The Concepts of Classical Thermodynamics}. Cambridge, Cambs: Cambridge University Press.

\bibitem{boyling} Boyling, J. B. (1968). Carathéodory’s principle and the existence of global integrating factors. \textit{Communications in Mathematical Physics}. 10(1): 52–68. doi.org/10.1007/BF01654133

\bibitem{botsko} Botsko, M. W. (2003). An Elementary Proof of Lebesgue’s Differentiation Theorem. \textit{The American Mathematical Monthly}. 110(9): 834–838. doi.org/10.2307/3647803

\bibitem{lavis} Lavis, D. A. (2019). The question of negative temperatures in thermodynamics and
statistical mechanics. \textit{Studies in History and Philosophy of Modern Physics}. 67: 26–63. doi.org/10.1016/j.shpsb.2019.02.002

\bibitem{lieb} Lieb, E. H., Yngvason, J. (1999). The physics and mathematics of the second law of thermodynamics. \textit{Physics Reports}. 310: 1–96. doi.org/	10.1016/S0370-1573(98)00082-9

\bibitem{nivaldo} Lemos, N. A. (2015). Vínculos dependentes de velocidades e condição de integrabilidade de Frobenius. \textit{Revista Brasileira de Ensino de Física}. 37(4). dx.doi.org/10.1590/S1806-11173731989

\end{thebibliography}
\end{document}